\documentstyle[12pt]{article}
\begin{document}
\title{Large N limit of O(N) vector models}
\author{Sigurd Schelstraete\thanks{Research Assistant NFWO, \ e-mail
address:\ schelstr@allserv.rug.ac.be}
\\  Henri Verschelde\thanks{Senior Reseach Associate NFWO,\ e-mail
address:\ hschelde@allserv.rug.ac.be} \\ \\
Department of Mathematical Physics and Astronomy \\
Universiteit Gent \\
Krijgslaan 281, S9  \\
9000 Gent, Belgium}
\maketitle
\begin{abstract}
Using a simple identity between various partial derivatives of the energy of
the vector model in 0+0 dimensions,
we derive explicit results for the coefficients of the
large N expansion of the model. These coefficients are functions
in a variable $\rho^2$, which is the expectation value of the two point
function in the limit $N=\infty$. These functions
are analytic and have only one (multiple)
pole in $\rho^2$. We show to all orders that these expressions obey
a given general formula. Using this
formula it is possible to derive the double scaling limit in an alternative
way. All the results obtained for the double scaling limit agree with earlier
calculations.
\end{abstract}
\newpage
\newcommand{\be}{\begin{equation}}
\newcommand{\ee}{\end{equation}}
\section{Introduction}
The $O(N)$ vector model has for some time been an interesting testing ground
for
ideas and techniques that are widely used in
the more complicated matrix-models. The latter
models recently gained importance as theories of two-dimensional gravity
coupled to matter with central charge $c \le 1$
\cite{doug,brez,gross}. In this article we investigate
the large-N expansion of the vectormodel, obtaining analytic expressions
for the various terms in this expansion. This proces can be easily continued
to any desired order, leading to exact closed expressions in every order. We
will
show that the results obtained here are in agreement with the well-known
perturbative expansions of vectormodels, as well as with the double-
scaling limit, considered earlier in \cite{Divec,zinn,moshe}. As an exta bonus
the method used
to obtain these expressions is useful to find non-perturbative
approximations for the energy of the vector model for any value of $N$.
The $O(N)$ vector model is defined by the following partition function:
\be
e^{-E} = \int d^N x e^{-\beta x^2 -  g (x^2)^2}
\ee
A crucial observation is now that for this model all correlation functions
(or, in this simple case,
the moments of the integral) can be obtained by successive derivations
with respect to the parameter $\beta$. We find for the first two correlators:
\be
{\partial E \over \partial \beta} = <x^2>
\ee
$${\partial ^2  E \over \partial \beta ^2} = <x^2>^2 - <(x^2)^2>
= ({\partial E \over \partial \beta})^2 - {\partial E \over \partial g} $$

We rewrite the last formula in a form that will be of greater use:
\be
\label{rel}
{\partial E \over \partial g} = ({\partial E \over \partial \beta})^2 -
{\partial ^2 E \over \partial \beta ^2}
\ee
It is indeed this formula that can be used to find a series expansion in
the coupling constant for any value of N, and will also form the basis
for the large N expansion of the model. Let us show how this comes about.
The only required input is the value of the energy for the
free model (i.e. without the interaction term). Since in this case the integral
is simply gaussian, this value is easily found to be:
$ E = {N \over 2} \log(\beta) + c^{te}$.
Using this expression as the lowest order approximation to the energy
one can calculate with it the RHS of (\ref{rel}). We then find in the LHS the
derivative
of the energy with respect to $g$, given up to zeroth order. Upon integration
we find the energy up to first order. Explicitly:
\be
E = {N \over 2} \log(\beta) + {{N(N+2)} \over {4 \beta ^2}} g + O(g^2)
\ee
This first order approximation to the energy
can again be used in the RHS of (\ref{rel}), giving us, after integration, the
energy to order $g^2$. It is clear that by just repeating this
simple algorithm the energy can be found
to arbitary large order in the coupling constant
for any value of $N$. The first few orders are given by:
$$E= {N \over 2} \log(\beta) + {{N(N+2)} \over {4 \beta^2}} g -
{N(N+2)(N+3) \over 4 \beta^2} g^2 $$
\be
+ {N(N+2)(5N^2+34N+60) \over 12 \beta^{12}}g^3
-{N(N+2)(7N^3+79N^2+310N+420) \over 8 \beta^8} g^4
\ee
Since the zero-dimensional integral simply counts the number of diagrams in
every order of perturbation, we explicitly find the number of diagrams
with one, two, three, ... index-loops in every order.

\section{Large N expansion}
The simple formula (\ref{rel}) can however be used to give far more powerfull
results than just a series expansion in the coupling constant
for general $N$. In fact, with a
little extra input it is all we need to find explicit expressions for
the coefficients of the large $N$ expansion of the integral.
The large N expansion of the vector model consists in taking
the number of field components $N$ to $\infty$,
at the same time rescaling the coupling constant $g$ to $g/N$. The $1/N$
expansion is then the series expansion
around $N=\infty$ of the model thus defined. It can equivalently
be viewed as an expansion in the number of indexloops of a given
diagram. For large $N$ the expansion can then be used as an approximation
to the original integral. It differs significantly from
the perturbation series in that every order provides us with a
nonperturbative approximation to the original function.
The general formula (\ref{rel}) needs a little modification in this case
because of the rescaling of g and is easily found to be:
\be
\label{rel2}
{\partial E \over \partial g} = {1 \over N}(({\partial E \over \partial
\beta})^2 -
{\partial ^2 E \over \partial \beta ^2})
\ee
The following expansion of the energy is proposed:
$E=N E_0 + E_1 + {1 \over N} E_2 + ....
= \sum_{i=0}^{\infty} N^{1-i} E_i$
Putting this into (\ref{rel2}) and identifying the terms with equal powers of
${1 \over N}$, we find an infinite series of identities:

$${\partial E_0 \over \partial g} = ({\partial E_0 \over \partial \beta})^2 $$

$${\partial E_1 \over \partial g} = 2({\partial E_0 \over \partial \beta})
({\partial E_1 \over \partial \beta}) - {\partial ^2 E_0 \over \partial \beta
^2}$$

\be
{\partial E_2 \over \partial g} = 2({\partial E_0 \over \partial \beta})
({\partial E_2 \over \partial \beta}) +({\partial E_1 \over \partial \beta})^2
-
{\partial ^2 E_1 \over \partial \beta ^2}
\ee
\noindent
and so on, with the p th formula given by:
\be
{\partial E_p \over \partial g} = \sum_{i=0}^{p}
({\partial E_i \over \partial \beta})
({\partial E_{p-i} \over \partial \beta})  -
{\partial ^2 E_{p-1} \over \partial \beta ^2}
\ee
We can rewrite this in a more convenient way as:
\be
\label{genform}
{\partial E_p \over \partial g} = 2({\partial E_0 \over \partial \beta})
({\partial E_p \over \partial \beta})+  \{
\sum_{i=1}^{p-1}({\partial E_i \over \partial \beta}) ({\partial E_{p-i} \over
\partial \beta})  -
{\partial ^2 E_{p-1} \over \partial \beta ^2} \}
\ee
In this form, the second part of the RHS of (\ref{genform})
consists entirely of lower order
coefficients in the large N expansion, while only the first term depends
on $E_p$. It is this property that will allow a recursive determination
of the coefficients.
We know that the energy is a dimensionless quantity, and that therefore
it can only depend on the dimensionless combination $g/\beta^2$. We thus find
the
following relation:
\be
\label{dim}
{\partial E_p \over \partial \beta} = -2 {g \over \beta}
{\partial E_p \over \partial g}
\ee
Using (\ref{dim}) in (\ref{genform}) we find:
\be
\label{Ep}
E_p = \int_{0}^g d \, g' {F_{p-1} \over 1+ {4 g' \over \beta} \rho^2  } \ ,
\ee
where $F_{p-1}$ is defined as being equal to the second term in the RHS
of (\ref{genform}) and $\rho^2$ is the expectation value
of  $x^2 \over N$ in the $N = \infty$ limit.
(i.e. the derivative of $E_0$ with respect to $\beta$).
Since $\rho^2$ and $E_0$ can be obtained from the saddle point approximation
of the integral, this formula will allow us to find
all the coefficients in the large N expansion using recursion.

\section{Coefficients of the large N expansion}
The saddle point equation of the integral is found by rescaling $x$ to
$\sqrt{N} x$ and finding the extrema of the effective action. This
gives rise to the following equation for $\rho^2$:
\be
\label{ro2}
1-2 \beta \rho^2 - 4 g \rho^4 = 0
\ee
for which we choose the regular solution (finite when $g=0$):

\be
\rho^2 = -{\beta \over 4 g} (1-\sqrt(1+{4 g \over \beta^2}))
\ee

Taking the derivative of (\ref{ro2}) with respect to $\beta$ and $g$
we easily find the following identities:
\be
{\partial \rho^2 \over \partial \beta} = - {\rho^2 \over \beta + 4 g \rho^2}
= - {\rho^4  \over 1-\beta \rho^2}
\ee
\be
\label{ro2g}
{\partial \rho^2 \over \partial g} =  {-2 \rho^4 \over \beta + 4 g \rho^2}
=  {-2 \rho^6  \over 1-\beta \rho^2}
\ee
Note that in the last expressions all explicit g-dependence is eliminated.
We now find an expression for $E_0$ by integrating the relation $\rho^2 =
{\partial E_0 \over \partial \beta}$, leading to:
\be
E_0 = {\beta \rho^2 - \log(\beta \rho^2) \over 2}
\ee
This is all one needs to obtain the coefficients of the large N expansion
to arbitrarely high order in $1 \over N$. Using formula (\ref{ro2g}), we can
change integration variables from $g$ to $\rho^2$. In doing so we completely
eliminate any explicit dependence on $g$. The expressions will be given
as functions of $\rho^2$ only. (or, more correctly, as functions of the
dimensionless quantity $\beta \rho^2$). After the change of variables
 formula (\ref{Ep}) becomes:

\be
E_p = -{\beta \over 2} \int_{1 \over {2 \beta}}^{\rho^2}
{F_{p-1}(\rho^2) \over \rho^4} d \rho^2
\ee

The integrals encountered in this problem are relatively simple and can all
be carried out exactly. Note that this would have been quite more complicated
without the change of variables. Since all lower order coefficients are
needed to find a coefficient of higher order, we have to determine the
expressions one by one in rising order. Beginning with $E_1$ we find
without too much difficulty:

$$F_0 = {\rho^4 \over 1-\beta \rho^2}  $$
$$E_1 = -{\beta \over 2} \int_{1 \over {2 \beta}}^{\rho^2}
{1 \over 1-\beta \rho^2}  $$
\be
= {1 \over 2} \ln(2-2 \beta \rho^2)
\ee
In higher orders the expressions become more elaborate, but no serious
complications arise. The integrals are always of the same type
(simple integrals of rational functions) and can
easily be carried out exactly. For $E_2$ we find in an analogous way:

$$F_1 = ({\partial E_1 \over \partial \beta})^2 -
{\partial ^2 E_1 \over \partial \beta ^2} $$
$$=-{\rho^4 \over 4}
{1+2 \beta \rho^2 -8 \beta ^2 \rho^4 \over (1-\beta \rho^2)^2}$$
\be
E_2=-{1 \over 24} {(2 \beta \rho^2 -1)^2 (\beta \rho^2 +4) \over (1-\beta
\rho^2)^3}
\ee

Explicit results for higher order are collected in an appendix. A closer look
at those results reveals that from $E_2$ on the coeffcients appear to obey
some general form:
\be
\label{gefor}
E_p = {(1-2 \beta \rho^2)^p \over (1-\beta \rho^2)^{3(p-1)}}
P_p(\beta \rho^2)
\ee
Where $P_p$ is a polynomial of degree at most $2p-3$. This form holds for
all the expressions that are given in the appendix. We will now prove
by induction that it is true for all orders.

We already know that the energy is a dimensionless quantity. It thus
depends only on the dimensonless variable $\beta \rho^2$, which we will
from now on denote as $\alpha$. The derivative of a function of $\alpha$
with respect to $\beta$ is given by

$${\partial \over \partial \beta} f(\alpha) = {\partial f \over \partial
\alpha}
{\partial \alpha \over \partial \beta}$$
\be
={\alpha \over \beta} {1-2\alpha \over 1-\alpha}
{\partial f \over \partial \alpha}
\ee

If the function $f$ happens to be of the form (\ref{gefor}),
the derivative with respect to $\beta$ is given by:

\be
{\partial f \over \partial \beta}={\alpha \over \beta}
{(2\alpha-1)^p \over (1-\alpha)^{3 p-1}} Q_p(\alpha)
\ee

With $Q_p$ a polynomial. The second derivative is given by:

\be
{\partial^2 f \over \partial \beta^2}={\alpha^2 \over \beta^2}
{(2\alpha-1)^p \over (1-\alpha)^{3 p+1}} Q^*_p(\alpha)
\ee

Now we assume that all $E_i$ obey the form (\ref{gefor}), i=1,...,p-1. In
that case $F_{p-1}$ is given by:

$$F_{p-1} = {\alpha^2 \over \beta^2}
{(2 \alpha-1)^{p-1} \over (1- \alpha)^{3 p-2} } Q^{**}_p(\alpha) $$
\be
=  \rho^4 {(2 \alpha-1)^{p-1} \over (1- \alpha)^{3 p-2}} Q^{**}_p(\alpha)
\ee

With this form for $F_{p-1}$, the coefficient $E_p$ is given by:
\be
\label{Ep2}
E_p = -{1 \over 2} \int_{1 \over 2}^{\alpha}
{(2\alpha '-1)^{p-1} \over (1-\alpha ')^{3 p-2}}
Q^{**}_{p}(\alpha ') d \alpha '
\ee
It is clear that also this integral is of the form proposed in (\ref{gefor}).
The integral can be performed by expanding the numerator in powers of
$1-\alpha '$. The integral is then reduced to a sum of simple integrals of
inverse powers of $(1-\alpha ')$, the highest power being $3p-2$. After
integration
the result is therefore still a sum of inverse powers of $1-\alpha$,
this time with highest power $3(p-1)$. The integral is is therefore given
by a polynomial divided by $(1-\alpha)^{3(p-1)}$. Furthermore, since the
integral vanishes for $\alpha={1 \over 2}$, the RHS must at least be
proportional
to $(2 \alpha-1)$. But, since the integrand is proportional
to $(2 \alpha ' -1)^{p-1}$,
the integral itself is proportional to $(2 \alpha-1)^p$. We have thus proved
by induction that the form proposed for the p th coefficient in the large
N expansion of the energy of the vector model is indeed correct. This
general form can be used to obtain the double scaling limit of the vector
model.
Two remarks are at order here: in the proof we assume that $E_1$ has the
form proposed in (\ref{gefor}). This is clearly not the case. However, the
derivative
of $E_1$ has the correct form, and therefore the proof remains
valid. Secondly, one sees that the general form bears a striking
ressemblance to the form conjectured for the coefficients of the large
$N$ expansion of matrix models. In \cite{bessis} the following ansatz
was proposed for these coefficients:
\be
E_h = {(1-a^2)^{2h-1} P_h(a^2) \over (2-a^2)^{5(h-1)}}
\ \ \ h \ge 2
\ee
\noindent
with $P_h$ a polynomial and $a^2$ the solution of a quadratic equation
similar to (\ref{ro2}).
\section{Double scaling limit}
{}From (\ref{gefor}) we find that all the coefficients exhibit a pole
at $\alpha = 1$. This
pole corresponds to the branch point singularity of $\rho^2$, namely:
$g_c=-{\beta^2 \over 4}$. If we define $\epsilon$ by $\epsilon \beta^2 =
g-g_c$,
$\alpha$ can be developed around its critical value and we find:
$\alpha = 1-2 \sqrt{\epsilon} + O(\epsilon)$.
Substituting this in the general form
we have found for the coefficients of the large $N$ expansion we find
that near the critical point they behave as:
\be
E_p \simeq ({1 \over 8})^{p-1} P_p(1) \epsilon^{-{3 \over 2} (p-1)}
\ee
This is of course not true for $E_0$ and $E_1$. It is however
straightforward to obtain their expansion around the critical point
with the expressions we have found:
$$E_0(g)-E_0(0) = {1 \over 4} -{\log(2) \over 2} +
\epsilon -{8 \over 3} \epsilon^{3 \over 2} +O(\epsilon^2)$$
\be
E_1 = {1 \over 2} \log(4 \sqrt{\epsilon}) + O(\epsilon)
\ee
This is exactly the same behaviour found in \cite{Divec} (because of a
different
definition of $\epsilon$, one has to replace $\epsilon$ in the above
formulae with $4 \epsilon$). The behaviour of $E_0$ and $E_1$ was dubbed
"non-universal" and therefore these terms were omitted in the double
scaling limit. We will follow this procedure and define the energy as
$E=\sum_{p=2}^{\infty} {1 \over N^{p-1}} E_p$. Near the critical point
the energy is approximated by:
\be
E \simeq \sum_{p=2}^{\infty} N^{-(p-1)} ({\epsilon^{-{3 \over 2}}})^{p-1}
P_p(1) ({1 \over 8})^{p-1}
\ee
One remarks that in every term we now have a competition between a
suppression factor proportional with an inverse power of $N$ and a
divergence proportional to a negative power of $\epsilon$. The double
scaling limit consists in taking $N$ to $\infty$ and $\epsilon$ to $0$,
while keeping the combination $N \epsilon^{3 \over 2}$ finite
(this is the same combination as found in \cite{Divec}). In this
way we sum the contributions from all loops. If we define the double scaling
variable as: $z= (N \epsilon^{3 \over 2})^{-1}$ the energy is given by
the series expansion: $E=\sum z^{p-1} ({1 \over 8})^{p-1} P_p(1)$.
We will now prove
that the complete series expansion is indeed the one proposed in \cite{Divec}.
We find that in the double scaling limit $(\alpha \simeq 1-2 \sqrt{\epsilon})$
the following relation holds:
\be
{\partial f(\alpha) \over \partial \beta} \simeq {-\alpha \over
{2 \sqrt{\epsilon} \beta}} {\partial f(\alpha) \over \partial \alpha}
\simeq {1 \over 2 \beta}  {\partial f(\alpha) \over \partial \epsilon}
\ee
Therefore the first and second derivatives with respect to $\beta$
are given by:
$$ {\partial E_p \over \partial \beta} \simeq {-3(p-1) \alpha
\over 4 \beta} b_{p-1} \epsilon^{-{3 \over 2} p + {1 \over 2}} $$
\be
{\partial^2 E_p \over \partial \beta^2} \simeq {3(p-1)(3p-1) \alpha^2
\over 16 \beta^2} b_{p-1} \epsilon^{-{3 \over 2} p - {1 \over 2}}
\ee
\noindent
(and similarly for $E_1$, where $E_p$ is assumed to have the form
$b_{p-1} \epsilon^{-{3 \over 2} (p-1)}$)
Another usefull relation is given by:
\be
F_{p-1} = 2 {\sqrt{\epsilon} \over \beta^2}
{\partial E_p \over \partial \epsilon} \ ,
\ee
which we find by changing variables in (\ref{Ep2}) from $\alpha$ to $\epsilon$
and taking the derivative of both sides with respect to $\epsilon$.
Using these relations in the definition of $F_{p-1}$ and identifying
equal powers of $\epsilon$ we readily find a recursion relation for
the coefficients $b_{p}$:
\be
\label{rec}
-4 p b_p = {3 \over 4} \sum_{q=2}^{p-1} b_{q-1} b_{p-q} (p-q)(q-1) -
{3 \over 4} p(p-1) b_{p-1}
\ee
Now, from the definition of the energy in the double scaling limit we find:
\be
E=\sum_{p=2}^{\infty} b_{p-1} z^{p-1} = \sum_{p=1}^{\infty} b_p z^p
\ee
Therefore the recursion relation (\ref{rec}) can be cast into a differential
equation for the energy. This can be done by multiplying the relation
with $z^{p-1}$ and summing p from 2 to $\infty$.
\be
-4 \sum_{p=2}^{\infty} p b_p z^{p-1} = {3 \over 4} \sum_{p=3}^{\infty}
\sum_{q=2}^{p-1} b_{q-1} b_{p-q} (p-q)(q-1) z^{p-1} -
{3 \over 4} \sum_{p=2}^{\infty} p(p-1) b_{p-1} z^{p-1}
\ee
\noindent
The LHS is immediately found to be equal to:
\be
-4 {\partial E \over \partial z} + 4 b_1
\ee
\noindent
While the first term in the RHS is equal to:
\be{3 \over 4} ({\sum_{q=2}^{\infty} b_{q-1} (q-1) z^{q-1}})^2 = {3 \over 4}
(z {\partial E \over \partial z})^2
\ee
\noindent
And the second term finally:
\be
-{3 \over 4} z {\partial^2 (z E) \over \partial z^2}
\ee
Defining $\xi$ as $\partial E \over \partial z$ this leads to the
following differential equation:
\be
3 z^2 {\partial \xi \over \partial z} -3 z^2 \xi ^2 + 6 z \xi
-16 \xi = 5/12
\ee
\noindent
One can also derive an equation for the partition function $\zeta = e^{-E}$:
\be
z^2 {\partial^2 \zeta \over \partial z^2} +2 z {\partial \zeta
\over \partial z} -{16 \over 3} {\partial \zeta \over \partial z}
+ {5 \over 36} \zeta =0
\ee
\noindent
This is exactly the same equation as found in \cite{Divec}, apart from a
rescaling
from z to $z/8$.
We have therefore proved that the double scaling limit defined in this
way confirms the earlier derivations.
\section{Conclusions}
In this paper we have explicitly obtained the large $N$ expansion of
the vector model. The coefficients are found to be analytic functions of
a variable $\rho^2$, which is the expectation value of the two point
function in the limit $N=\infty$. Remarkably enough all these functions
(apart from the first two) obey some general form, which clearly
illustrates the divergence of the coefficients. It is seen that all
coefficients diverge at the same critical value of the coupling, and
in such a way that a double scaling limit may be taken. This alternative
derivation is in a sense closer to the philosophy of the double scaling
limit because it clearly illustrates the competition between the large
$N$ suppression and the divergence at critical coupling. The results
derived in this paper are however far more general than their application
to the double scaling limit. They also allow us to obtain information
of the model away from the critical region and to find approximations
for the energy for finite values of $N$.

\appendix
\section{Appendix}
We collect some important results in this appendix. The first six
coefficients in the large $N$ expansion are given by:

$$E_0 = {\beta \rho^2 - \log(\beta \rho^2) \over 2}$$
$$E_1= {1 \over 2} \ln(2-2 \beta \rho^2) $$
$$E_2=-{1 \over 24} {(2 \beta \rho^2 -1)^2 (\beta \rho^2 +4) \over (1-\beta
\rho^2)^3}$$
$$E_3=-{5 \over 16} {(2 \beta \rho^2 -1)^3 \beta \rho^2 \over (1-\beta
\rho^2)^6}$$
$$E_4 = {(2 \beta \rho ^2 -1)^4 (8 \beta^5 \rho^{10} -56 \beta^4 \rho^8
+164 \beta^3 \rho^6 -4696 \beta^2 \rho^4 -1841 \beta \rho^2 +896) \over
5760 (1-\beta \rho^2)^9}$$
$$E_5 =- {(2 \beta \rho^2 -1)^5 \beta \rho^2 (706 \beta^2 \rho^4 +
823 \beta \rho^2 -399) \over 256 (1-\beta \rho^2)^{12}} $$
\noindent
The first terms of the series expansions of these expressions are
given by:
\begin{eqnarray*}
E_0 &=& g -4g^2 +{80\over 3} g^3 -224g^4 +{1072 \over 5} g^5 +...  \\
E_1 &=& 2g -20g^2 +{704 \over 3} g^3 -2976 g^4 +{197632 \over 5} g^5 +... \\
E_2 &=&  -24g^2 + {2048 \over 3} g^3 -14976 g^4 + 296960 g^5 +... \\
E_3 &=& 640 g^3 - 33280g^4 +1126400 g^5 +... \\
E_4 &=& -26880 g^4 + {10604544 \over 5} g^5 + ... \\
E_5 &=& 1548288 g^5 + ...
\end{eqnarray*}
\noindent
Notice that the sum $\sum_{i} E_i$ gives the correct expansion of the
"$N=1$ vectormodel", which is just a one-dimensional integral:
$$E=3g -48g^2 +1584 g^3 -78336g^4+{25671168 \over 5} g^5 +...  $$
\noindent
(where we have put $\beta = {1 \over 2}$)

\end{document}